# “What Is It That You Don’t Understand?” Language Games and Black Box Algorithms*

Rémy Demichelis
University of Turin

*Abstract*
The aim of this article is to understand the problem of “black box” algorithms, an issue inherent to the nascent field of Explainable Artificial Intelligence (XAI). While it is relatively easy to understand something someone explained to us, it becomes more complicated when no one can fully grasp the issue. Our purpose is however to highlight: (1) that we should speak of *interpretability* rather than *explainability* when we seek to understand models, mainly because we never have complete and unambiguous access to information; (2) that the machines face the problem of the inscrutability of reference, in the same way that the linguist imagined by Willard Van Orman Quine cannot precisely determine what the term “*gavagai*” refers to in a situation of radical translation; (3) that there is no rule for the application of language, except for “language games”, as Ludwig Wittgenstein’s linguistics teaches us. The hope of achieving complete explicability and transparency of algorithms is undoubtedly in vain: we can only rely on partial and broad interpretations that will never fully explain the underlying rules.

## 1. A school memory

Which one was it? I no longer know. There were certainly several of them, those teachers who asked me: “What is it that you don’t understand?” A sentence heard so often that it has become woven into the collective imagination of France’s public education system, in a blend of kindness and terror. How can one answer this seemingly innocent question? To understand what I don’t understand, I would first need a clue, a lead, some inkling of where I should be heading. In short, I would need to have understood a little in order to know what it is that I don’t understand. But shame arises from the very inability to put your ignorance into words; you find yourself stupider than stupid.

Today, we are prey to this same embarrassment when we try to explain certain artificial intelligence (AI) algorithms, because their reasoning eludes mathematical formulation. This is not entirely by chance, if we believe Bruno Bachimont's definition:

> AI aims to process computationally (that is the method) problems that require non-formalisable knowledge in order to be solved (that is the object). (Bachimont, 1994, p. 181)

If this knowledge cannot be formalised, then trying to formalise an explanation of it—therefore a form of knowledge—is to embark on a paradoxical task. Indeed, the least explainable systems (machine-learning systems based on statistical inferences, and especially artificial neural networks or deep learning) have proved the most useful for purposes as varied as image recognition, translation, or content generation. By definition, these algorithms no longer lend themselves to a formal, mathematical-type logical formulation, which gives them greater flexibility and therefore adaptability, but they thereby escape our understanding.

It has become common to speak of black boxes, in the sense that the results produced are not easily explained. At best, we can rely on estimates using other statistical methods, but not on the kind of explainability that mathematics provides. We therefore consider that it is more appropriate to speak of interpretability when we cannot achieve unambiguity. However, optimistic minds do not let up in their efforts and still hope to reach explainability, to the point of giving rise to the new discipline of Explainable Artificial Intelligence (XAI).

But before any rescue attempt, the crucial question should be: why exactly are we unable to achieve this absence of ambiguity? In short, let us confront our childhood demons and ask ourselves: what is it that I don't understand?

Our aim is to show that the answer is certainly to be sought on the side of linguistics, in the sense that language does not allow us to state the rules that enable us to name things—and that the same holds for AI systems based on deep learning. We place ourselves among the pessimists.

To support this thesis, we will begin by detailing how deep-learning systems work, in order to better grasp the issue. We will then review the state of the art in XAI. Next, we will look at the difficulty of identifying the criterion for naming within the philosophical tradition, mainly through Wittgenstein. Finally, we will draw on his argument to develop our own thesis.

## 2. What is the problem?

When faced with the question "why did the AI system produce this result rather than another?", it has become difficult to answer since the revolution brought about by machine learning using artificial neural networks, so-called "connectionist" AI. Up until the 2010s, "symbolic" AI dominated the field of computer science (Mitchell, 2019, Location 366), and the problem of explainability did not really arise: operations were applied to symbols that meant something to us (red, dog, big, etc.). Then, a few innovations (Krizhevsky et al., 2012; Le et al., 2012) in computer vision changed the paradigm and restored artificial neural networks to prominence. They were mainly supported by the huge amounts of data newly available thanks to the development of the Internet. Such masses of data are necessary for machine learning, because artificial neural networks need large quantities of information to train and automatically infer the best parameter settings to solve a problem.

However, this means that the system remains strongly influenced by its training data, without our being able to identify where or how this influence is exerted within the algorithm. In this way, an image-categorization system may mistakenly come to equate snowy landscapes with photos of wolves (Ribeiro et al., 2016), because all the instances of wolves in its training database showed the canids in snow. In the same way, a system may start systematically adding a human arm to a dumbbell when it is asked simply for “a dumbbell”, because all the dumbbells in the training photos were being held at arm’s length (Mordvintsev et al., 2015). We often say that there is a bias in this kind of situation. The underlying problem is that associations of ideas or automated expectations can then be created between people and facts or behaviors. Thus, a multitude of discriminatory biases have been observed in AI systems, reproducing sexist, racist, Islamophobic, or ableist prejudices (Broussard, 2023; Buolamwini, 2023; Demichelis, 2024a).

Identifying these pitfalls, isolating them, and countering them has become a top-priority issue in our democracies. The European regulation known as the “AI Act”, published in 2024, provides that so-called “high-risk” systems, used in particular for human resources or credit allocation, must be audited in order to identify “possible biases that are likely to affect the health and safety of persons, have a negative impact on fundamental rights or lead to discrimination prohibited under Union law, especially where data outputs influence inputs for future operations”. (Regulation 2024/1689, 2024, Articles 10–2, f) Publishers will therefore have to take “appropriate measures to detect, prevent and mitigate possible biases”. (Regulation 2024/1689, 2024, Articles 10–2, g)

The great difficulty for computer scientists is to succeed in understanding how deep-learning algorithms work, because the value of each parameter is rarely intelligible. In the lower layers, connectionist systems do not manipulate symbols, such as a color code like #FF0000 for red, but numerical values derived from that code and transformed through multiple operations. These calculations are sometimes very simple, but because they are performed in a cross-connected way, the reference is diluted as the calculations proceed: at neuron 456, it is impossible to know what the number 0.42 means. It means nothing, much like the super-powerful computer in the film *The Hitchhiker’s Guide to the Galaxy*, which answers “the question of life, the universe and everything” with “42” (Jennings, 2005, Location 40:35:00). It makes no sense and leaves us deeply embarrassed; we do not know what to do with this information. The problem is even greater because, from the vectorization stage (when the symbol is transformed into numerical values), the symbols of our natural language often disappear; in language processing, the software rarely considers a word in isolation when incorporating it into its calculations, but rather as part of a group of several words, within a certain context. As a result, the vectors to which operations are applied represent only a set of words or letters that lose their meaning for us.

## 3. Solutions and Their Limits

Although it is difficult to identify exactly which parameters, vectors, or data were decisive in an AI system’s decision, this has not prevented many researchers from trying to take up the challenge. The explainability of algorithms has thus become a fully-fledged branch of computer-science research known as Explainable AI, or XAI (Saeed & Omlin, 2023). The goal, then, is to achieve *explainability*; however, we argue (as professionals also shared with us during [informal talks]) that what is really at stake is rather *interpretability*, because it is impossible to reach mathematical unambiguity (what the term explainability refers to) with opaque statistical systems.

Two approaches are often mentioned in the literature when discussing different types of interpretation: the *local* approach and the *global* approach (John-Mathews, 2019). The local approach consists in identifying the determinants for a given individual: why was a particular person’s loan application rejected? Why was a particular dog correctly categorized as a dog? Etc. The global approach seeks to describe how the model works in general, regardless of individuals: what are the determining factors in animal classification? What information is crucial to obtain a loan? Etc. The two methods are not mutually exclusive and can be used in a complementary way. The local approach, however, tends to be applied to problematic use cases, where one seeks to understand why an individual was rejected or incorrectly categorized.

There is a plethora of technical solutions for interpreting an algorithm, but two are regularly cited: LIME (Ribeiro et al., 2016) and SHAP (Lundberg & Lee, 2017). Both aim to overcome the pitfall of relying only on standardized tests to measure model accuracy. An algorithm may perform very well under laboratory conditions (standardized tests) but not “in the wild” (when faced with examples or combinations it has never encountered and that are closer to real life).

LIME is designed to produce local interpretations for any type of algorithm. Its method consists in measuring the weight of certain input information: which words made it possible to say that a text was about Christianity or atheism? Which pixels made it possible to say that the image showed a husky rather than a wolf? LIME does not provide a global interpretation, but rather multiple local interpretations across different outputs to offer a better understanding of the model.

SHAP, for its part, aims to estimate the weight of each input feature (like LIME), but no longer in a binary way (depending on their presence or absence); instead, it does so by using values (Shapley values). The aim is to measure robustness to contextual variability and to ensure the consistency of the interpretation, notably if a decisive feature sees its weight increase or remain the same (consistency).

It is interesting to note that the papers on LIME and SHAP both rely on humans to evaluate the intelligibility of interpretations. “Explanations should be easy to understand” and aim at “a qualitative understanding”, the first states, while the second mentions “human intuition” to validate its methodology. Understanding is a qualitative aspect that cannot be formalized and that requires an exchange with others to know whether they have understood. Only their subjective sense of understanding can serve as evidence.

For this reason, it is often suggested that users and the people targeted by a technology be involved both in auditing (Broussard, 2023, p. 163) and in building models (Achiume & UN. Human Rights Council, 2020). They are sometimes best placed to identify pitfalls and inconsistencies, refine the interpretation, or point out a lack of interpretability in the tool, depending on the situation. Even if a model is explainable, it may not be explainable for a particular community: one does not address a scientist in the same way as a literary scholar. It is therefore necessary to adopt a hermeneutic approach to interpretation. Behind this pleonasm lies a reference to hermeneutics as social critique; an interpretation grounded in the norms and knowledge (Regulation 2024/1689, 2024, Articles 5, c; Ribeiro et al., 2016) specific to a community and as close to it as possible (Graziani et al., 2022), rather than from an external standpoint that would deliver an top-down explanation like an imperial judge.

Olya Kudina speaks of "Interpretative Phenomenological Analysis", insofar as this method focuses on "situated micro-perspectives and on the philosophical principles of circular interpretation" (Kudina, 2023, p. 12); technology both brings a society's values to the fore and influences them, prompting their reactivation in a movement of "value dynamism" (Kudina, 2023, p. 3), or dialectic. To obtain a better understanding of the model, iterative work is needed; work that can take user feedback into account as usage unfolds.

Fabio Paglieri, for his part, suggests "following the money" in order to better understand how tools are steered:

> The explanations sought after in XAI (...) always concern the inner workings of AI systems, "how the magic happens", which is exactly what makes this quest so elusive for generative AI and ML. It is somewhat surprising, however, that little or no attention is given to other types of explanations, focused not on how these systems work, but rather on who stands to gain (or lose) from the fact that they do work. "Cui prodest?" (Paglieri, 2024, p. 55)

There is, in this idea, a certain *hermeneutics of suspicion* (Johann, 2020, p. 158). That is, an interpretation that does not hesitate to speculate about a speaker's less avowable and sometimes unconscious motives. The hermeneutics of suspicion finds its form in psychoanalysis or Marxism, insofar as there is often more to be read in the words of a patient or a political opponent than what is explicitly expressed. The idea is that something is hidden (intentionally or not) and that it must be brought to light. It is a delicate art that can tip into a kind of madness, into eccentric overinterpretations, into pathologies of interpretation. In the political, economic, and technological spheres, however, the hermeneutics of suspicion is also a way of not taking the statements of major digital companies at face value and of thwarting obfuscation strategies. Interpretation becomes demystifying. "Following the money" makes it possible to highlight the reasons that drive the production of tools and to understand why they are configured one way rather than another. When there are budget cuts for experimentation, ethics, or compliance, we better understand why certain systems break the law or violate moral norms. We better understand why certain content is promoted rather than other content when the sole objective is to keep internet users on a website in order to satisfy advertisers and a business model based on advertising. The examples are illimited.

It is interesting to note that all these interpretability methods more or less take for granted that the original model cannot be perfectly explained. As Marco Ribeiro et al. (in the paper on LIME) write: "it is often impossible for an explanation to be completely faithful unless it is the complete description of the model itself." The software and the various methods discussed in this section therefore do not truly aim at *explainability*, but at *interpretability*. There will always remain some uncertainty about cause-and-effect relationships in the original model. Abandoning the ambition of explainability also means giving up on total transparency. We will not be able to read an artificial neural network like an open book (and even a book requires interpretation). From this point of view, only *interpretability* methods will be *satisfactory*, but they will be *unsatisfactory* if the goal is the crystal-clear transparency promised by *explainability*.

## 4. PHILOSOPHICAL ASPECTS: THE NAME, THE MAP, AND THE GLASS

The difficulty of clearly identifying what, in an image, defines a cat, a wolf, or any other object was already expressed in a different way in Antiquity. To mock Plato, who had defined man as a "featherless biped", Diogenes of Sinope (known as the Cynic) brought him a plucked rooster and exclaimed: "Here is Plato's man!" (Laërtius, 1925, Book VI, 40) This anecdote-parable shows that any definition always risks omitting certain aspects of an object. And what about accidents? A man without legs is no longer a biped and yet is no less a man. Thus, listing criteria in order to categorize objects risks not only neglecting certain aspects, but also overlooking exceptions.

The idea that the explanatory model (the definition) is not the original model (the object) appears in scientific literature long before AI: "The map is not the territory", Alfred Korzybski said in 1931 (Korzybski, 2004). Legend has it that this phrase was inspired by a grim experience: during the First World War, soldiers under his command were allegedly killed by a Prussian machine gun that was not shown on the map (Krivine, 2018, p. 98).

But more than that: the explanatory model must not be the original model. In 1946, Jorge Luis Borges (2004) wrote an absurd short story, *On Exactitude in Science*, in which cartographers create a map that represents exactly the entire territory of an Empire—that is, at a 1:1 scale. However, later generations quite naturally deem it "useless" (*inútil*). This means that representing the Empire symmetrically, "point for point", is of no use, and that the usefulness of a map lies precisely in synthesizing information, summarizing. A map must contain less information than the territory it represents to be of any use. With AI, the explanatory model must contain less information than the original model, even if the original model is explainable, that is, unambiguous. As Ribeiro et al. write:

> If hundreds or thousands of features significantly contribute to a prediction, it is not reasonable to expect any user to comprehend why the prediction was made, even if individual weights can be inspected. (Ribeiro et al., 2016)

The metaphor of the uselessness of the *perfect map* also conveys the idea that there is no *perfect term* that can correspond to the thing it refers to. If we think of a word as a map, then it cannot cover its referent perfectly. Like maps, concepts involve mediation, translation, and almost a betrayal of the thing aimed at, in short, an interpretation. This aligns with what we said about the vanity of any project of explainability and about the necessary mourning of total transparency. Emmanuel Alloa notes that a "a glass that is truly transparent ends up as negating its very materialexistence" (Alloa, 2022); there is transparency only because there is first an obstacle. He adds: "To promise free and informed circulation, ends up confining movement to a meticulously predesigned pattern.”

Let us apply this to AI: assigning a predetermined path to an artificial neural network would amount to limiting its potential. Deep learning benefits from its flexibility to adapt to varied situations for which rigid rules are insufficient. If we had to freeze rigid rules into the algorithm in the hope of making it explainable, we would strip it of its adaptability and its original usefulness; a good old-fashioned AI could do just as well, and artificial neural networks would therefore become obsolete. Even so-called “hybrid/neuro-symbolic” (Braunschweig, 2024; Trinh et al., 2024) or “reasoning” AIs (Brown et al., 2020; Leveau-Vallier, 2024) largely retain the flexibility granted by statistical inference.

Diogenes, Borges, Korzybski, and Alloa show us that concepts, maps, or modeling never achieve the goal of perfectly corresponding to the thing they are supposed to represent. We grasp that it is in the very nature of the medium used to contain a practical dimension, sometimes portable, in order to be applicable. In XAI, the sought-after application is understanding, but it cannot come through an attempt at a “point for point” description. In this way, AI contributes several centuries late to the medieval quarrel over universals (Conti, 2009): on the one hand, the realists believed in an isomorphism between the world, things, and words; on the other, the nominalists saw our categories only as semantic objects rather than real essences. Connectionist AI seems to argue for the latter and adds a certain dose of skepticism. There will be no definitive symbolic model of essences; there will only be rough estimates (without this being pejorative) that remain intimately dependent on the modalities of access.

However, is this the only limit to our understanding of computational models? So far, we have approached the question of interpretability from the angle of what is actively described. But something else is at play in our knowledge. We must now address the issue through the learning process in a situated way.

For a concept, unlike a map, rarely applies to a single particular case, to a single territory, but to several, whose specificities feed back into meaning. The meaning of words depends intimately on their uses, across different contexts and situations. In other words, every application enriches meaning and thus gives rise to generalized learning from situated configurations.

## 5. The Deep Linguistic Problem

The difficulty that engineers face in understanding what the machine truly designates is similar what encountered the “linguist” described by Willard Van Orman Quine (1960, Chapter 2). The philosopher imagines a situation of “radical translation”; a situation in which two people, each speaking a different language, meet and attempt to exchange words for the first time. With an ethnocentrism typical of his era, Quine envisions a dialogue between a “native” from an unknown land and the aforementioned linguist, who happens, of course, to be an English speaker.

If the native points to a rabbit and utters the term “*gavagai*”, what conclusion should the linguist draw? “*Gavagai*” could just as easily designate the rabbit, its ears, its whiskers, its entire head, or even the animal in a specific position. There is an “inscrutability of reference”, as the history of philosophy calls this problem [1], or an “indeterminacy of translation”, as Quine writes (1969, p. 35).

Without being able to rely on other concepts, the linguist is at a loss. He must resort to an inference to the best possible interpretation based on context, much like a deep learning system. We should therefore not be surprised to discover that the term “wolf” designates “snow” for a computer vision system: it finds itself facing the inscrutability of reference and attempts to overcome it as best it can. Without other elements to indicate otherwise, it may continue to believe for a long time that the snow in the image is called “wolf”.

The inscrutability of reference had already been highlighted by Ludwig Wittgenstein a few years earlier. He did not say “*gavagai*” though, but “tove” (a word that exists no more than Quine’s) (Wittgenstein, 1965, p. 2) to designate either a *pencil*, *roundness*, *wood*, *one*, *hard*, or something else entirely. And he writes that “it is the ostensive definition’s to *give* it a meaning” (Wittgenstein, 1965, p. 2).

For Wittgenstein, the meaning of a word is the fruit of “language games” (Wittgenstein, 2009, paras. 2 & 21). That is, meaning is constituted during situations in which a term is used ostensively; when someone points to something he/she is designating within a certain context. Wittgenstein uses the example of a worker showing materials or tools to another among various objects. The meaning of words thus emerges through their use.

> Language games are the forms of language with which a child begins to make use of words. The study of language games is the study of primitive forms of language or primitive languages. (Wittgenstein, 1965, p. 17)

In our view, deep learning AI systems – particularly, but not exclusively, computer vision – find themselves in similar situations during their training. What must be understood implicitly, and what Wittgenstein explains very well, is that there is no rule for the application of language as a set of rules; there is no rule for the application of the rule. Consequently, seeking to know why an AI system calls one object a “cat” and another a “dog” seems doomed to failure. We can list the characteristics of these species, but the way we apply concepts, or the way the machine does so, is a matter of usage rather than a rule. Usages are varied and depend on the context, like so many “games” in which language is used.

Humans have the flaw, according to Wittgenstein, of despising the particular in favor of the general, but it is this “craving for generality” (Wittgenstein, 1965, p. 18) that makes us lose sight of how meaning is formed; instead of looking at particular examples, because they are considered “incomplete” (Wittgenstein, 1965, p. 19), we pursue the generality of a rule that does not exist or that will always be deceptive, insufficient, and ultimately incomplete. Above all, it leads to an infinite regress, for once a rule is established, it still requires an “interpretation” (Wittgenstein, 1965, p. 33) to proceed with its application – an interpretation which can only consist of other rules, and so on. Regarding words, “we can't tabulate strict rules for their use" (Wittgenstein, 1965, p. 28).

[1] Quine did use the term “inscrutability” (Quine, 1960, p. 53) but not the concept “inscrutability of reference” to the best of our knowledge.

We ask AI systems to use language by emancipating themselves from rules. This is the reason why deep learning was used: because the enumeration of rules was not practicable, if not impossible. Computer scientists certainly turned to statistical inference out of practical necessity, because they were hitting a glass ceiling with symbolic AI (meaning they did not necessarily understand the difficulty they were facing), but in doing so, they illustrated Wittgenstein's point and, in a way, proved him right at the same time: *our use of language is a black box*.

Technological progress has been made possible by a statistical and ostensive approach to reference, whether through supervised, reinforcement, or unsupervised learning. If we show a succession of objects to a machine and assign them labels (supervised learning), we are using an ostensive approach *par excellence*. If we correct the machine based on the answers it provides (reinforcement learning), we are implicitly steering it toward the label we expect. If we let it search for categories on its own within data structured in such a way that its differentiation makes sense to us (unsupervised learning), we are still implicitly guiding the machine, albeit without acting on it directly. Regardless of the type of learning, we are showing it what we mean. Furthermore, if the structure of its responses strayed too far from our meanings, the machine would be immediately disqualified as being random and/or producing unintelligible remarks (it should be noted that the "hallucinations" of generative AI still mean something nonetheless).

Consequently, we may attempt to estimate the weight of each pixel or each word in the machine's parameters, but this will not allow us to understand exactly how it makes use of a word. Exactitude is not of this world, and that is precisely what makes our natural language so practical for expressing what we mean. It is also this lack of formalism, this open door to ambiguity, that provides artificial neural networks with capabilities never achieved before.

However, seeking to use rules to understand something that has no business being understood through rules is nonsensical. The rules we are looking for are not rules of application or comprehension, but of language itself: *models of explanation are merely language models*. *There is no rule of the rule*. If we now want to understand why neuron *x* gives result *y* and what its influence is on decision *z*, this question makes no more sense. If we want to know why the system provides a particular answer, the secret is simply that this is what we asked it to do by showing it examples that are not expressed as rules. At no point did we ask it to choose the most elegant route possible to get there; on the contrary, we asked for the most statistically efficient one. Why, then, should we be surprised by the lack of explainability?

There are, in fact, three problems within that of interpretability: (1) that of the weight of input information, which automatically excludes any explanation free of ambiguity; then (2) that of the system's functioning; and (3) that of application. We can estimate the weight of parameters without, however, arriving at a complete explanation; that is never the purpose of interpretability. There may be some clarification of the reasons, even economic or societal, that can be read into a particular decision, but a degree of obscurity will always remain. We can conversely have a complete view of the calculations without, however, achieving a comprehension of these operations. In any case, interpretations will not enlighten us regarding the rule of application, which vanishes and exists only through usage.

“The meaning of a phrase for us is characterized by the use we make of it” writes Wittgenstein (1965, p. 65). If we want to know more about a phrase, we can either study its cases of use during ostensive definitional processes, but this eliminates the possibility of finding the rule of application. Browsing through the training database to offer a qualitative analysis of it will prove useful without ever being sufficient. Or, we can analyze how the phrase fits into the language, and this is, in fact, the kind of research that interpretability systems propose. In other words, if we are looking for explainability, we will always be disappointed. Interpretation is a makeshift solution, but that does not mean it is useless.

## 6. Conclusion

Today, AI is faced with a problem for which practical answers are becoming necessary. This is the problem of “black box” algorithms; it has become a legal requirement to attempt to explain certain “high-risk” systems before their deployment. We have seen that several solutions exist to this end, whether technical or hermeneutic. These involve weighting parameters in a statistically quantifiable way, or seeking to understand the deeper reasons behind results, whether they are rooted in culture or in political and economic incentives.

However, these methods are never purely explanatory. They never offer the total absence of ambiguity required by mathematics. The only hope we have of understanding AI systems, particularly connectionist algorithms, a little better is to resort to *interpretation*. It is therefore more appropriate to speak of *interpretability* instead of XAI (Explainable AI). Yet, we did not stop at this observation; we also attempted to delineate the conditions that make *explainability* impossible.

First, we explored the idea that connectionist *AI systems have no ambition to be explainable at any point*, and that they actually free themselves from this constraint to achieve higher performance. Drawing on philosophical tradition, we then endeavored to provide a response already outlined by computer science researchers, which we can summarize as the idea that *so-called explanatory modeling can never perfectly correspond to the original model*. With the third response, we hope to have provided a more original approach. It suggests that *the process of learning a language relies on usages whose rules elude us*. It is not an impossibility of understanding the language itself, but an impossibility of formulating its usage outside the sphere of language; we understand, but manipulable symbols do not help us understand why we understand. For there is, via artificial neural networks or via our own minds, a residual inscrutability of reference when we speak of any object whatsoever.

This by no means implies that our brains and computer systems are identical. Of course, AI is built on the hope of reproducing our cognitive faculties and may present similarities, but there is no reason to take the leap into anthropomorphism. Even with a more or less analogous architecture, software may very well develop in a different manner and produce the same results. If engineers seem to prove Wittgenstein right regarding how language learning functions, they prove nothing else. To say that the machine "understands" would even be going beyond our point.

This critique of the ambition for absolute transparency must not however make us forget that there are compliance and moral stakes in attempting to sketch an interpretation of these tools. In a liberal society, it would be unacceptable for this effort not to be made when we know that these systems lead to harmful discrimination against certain categories of the population – people who are already susceptible to suffering injustices today. Insisting on the limits of interpretability must not lead to stagnation. Highly theoretical discussions too often serve to conceal what is merely bad faith, laziness, or even complicity.

To better understand algorithms, we must literally think "outside the box". We must move away from the heart of the machine and accept that its results certainly have more to tell us than its mechanisms, but that neither will tell us everything. We are not the first to suggest a "step to the side" in the methodology of interpretability, statistical or hermeneutic approaches are already prime examples of this.